\documentclass{article}%
\usepackage{amsfonts}
\usepackage{amsmath}
\usepackage{amssymb}
\usepackage{graphicx}%
\newtheorem{theorem}{Theorem}

\newtheorem{proposition}[theorem]{Proposition}
\newtheorem{remark}[theorem]{Remark}

\newenvironment{proof}[1][Proof]{\noindent\textbf{#1.} }{\ \rule{0.5em}{0.5em}}
\newcommand{\bpartial}{\mathop{\partial\kern -4pt\raisebox{.8pt}{$|$}}}
\newcommand{\bra}{\mathopen{[\kern-1.6pt[}}
\newcommand{\ket}{\mathclose{]\kern-1.5pt]}}
\newcommand{\bbra}{\mathopen{[\kern-2.2pt[\kern-2.3pt[}}
\newcommand{\bket}{\mathclose{]\kern-2.1pt]\kern-2.3pt]}}

\newcommand{\slx}{\mbox{\bfseries\slshape x}}
\newcommand{\sly}{\mbox{\bfseries\slshape y}}

\begin{document}

\title{A Maxwell like Formulation of Gravitational Theory in Minkowski
Spacetime\thanks{This is a version of a paper published in \textit{Int. J.
Mod. Phs. D} \textbf{16}(6), 1027-1041 (2007) where some misprints and typos
have been corrected, some references have been updated, a footnote has been
added and some few sentences have been rewritten to better explain the role of
the (plastic) deformation tensor $h$.}}
\author{E. A. Notte-Cuello$^{(1)}$ and W. A. Rodrigues Jr.$^{(2)}$\\$^{(1)}${\small Departamento de Matem\'{a}ticas,Universidad de La Serena}\\{\small Av. Cisternas 1200, La Serena-Chile}\\$^{(2)}${\small \ Institute of Mathematics, Statistics and Scientific
Computation}\\{\small IMECC-UNICAMP CP 6065 \ 13083-859 Campinas, SP, Brazil}\\{\small walrod@ime.unicamp.br; enotte@userena.cl}}
\date{22 June 2009}
\maketitle

\begin{abstract}
In this paper using the Clifford bundle formalism a Lagrangian theory of the
Yang-Mills type (with a gauge fixing term and an auto interacting term) for
the gravitational field in Minkowski spacetime is presented. It is shown how
two simple hypothesis permits the interpretation of the formalism in terms of
effective Lorentzian or teleparallel geometries. In the case of a Lorentzian
geometry interpretation of the theory the filed equations are shown to be
equivalent to Einstein's equations.

\end{abstract}

\section{Introduction}

In this paper we present a Lagrangian theory of the gravitational field in
Minkowski spacetime\footnote{Minkowski spacetime will be called Lorentz
vacuum, in what follows. Moreover in the pentuple $(M\simeq\mathbb{R}^{4},%
\mbox{\boldmath{$\eta$}}%
,D\mathbf{,\tau}_{%
\mbox{\boldmath{$\eta$}}%
},\mathbf{\uparrow)}$, $%
\mbox{\boldmath{$\eta$}}%
$ is a Minkowski metric, $D$ is its Levi-Civita connection, $\mathbf{\tau}_{%
\mbox{\boldmath{$\eta$}}%
}$ is the volume element defining a global orientation and $\uparrow$ refers
to a time orientability. The objects in the Loretnzian spacetime structure
$(M\simeq\mathbb{R}^{4},\mathtt{\mathbf{g}},\nabla,\tau_{\mathtt{\mathbf{g}}%
},\uparrow)$\ have analogous meanings. Details are given in,
e.g.,\cite{sawu,rodoliv2006}.} $(M\simeq\mathbb{R}^{4},%
\mbox{\boldmath{$\eta$}}%
,D\mathbf{,\tau}_{%
\mbox{\boldmath{$\eta$}}%
},\mathbf{\uparrow}$) which is of the Yang-Mills type (containing moreover a
gauge fixing term and an auto interaction term related to the vorticity of the
fields). In our theory each nontrivial gravitational field configuration can
be interpreted as generating an effective Lorentzian spacetime $(M\simeq
\mathbb{R}^{4},\mathtt{\mathbf{g}},\nabla,\tau_{\mathtt{\mathbf{g}}}%
,\uparrow)$ where $\mathtt{\mathbf{g}}$ satisfies Einstein equations or by an
effective teleparallel spacetime. Our theory is invariant under
diffeomorphisms and under local Lorentz transformations, and is based on
\textit{two} assumptions. The first one is that the gravitational field is a
set of Maxwell like fields, which are physical fields in Faraday sense (i.e.,
of the same ontology as the electromagnetic field, having nothing a priory to
do with the geometry of spacetime), which lives in Minkowski spacetime, have
\ its dynamics described by a specified Lagrangian density and couples
universally with the matter fields. Such coupling is such that the presence of
energy-momentum due to matter fields in some region of Minkowski spacetime
distorts the Lorentz vacuum in much the same way that stresses in an elastic
body produces deformations in it. This \textit{distortion} permits the
introduction of a new metric field $\mathtt{\mathbf{g}}$ in $M$ which is the
analogous of the Cauchy-Green tensor \cite{frankel} of elasticity theory. The
field $\mathtt{\mathbf{g}}$ can be written in terms of the Maxwell like fields
(potentials) describing the gravitational field in an appropriate way. Once
the Levi-Civita connection $\nabla$ of $\mathtt{\mathbf{g}}$ is introduced in
the game, it is possible to show that the Maxwell like field equations for the
gravitational fields which follows from the variational field implies that
$\mathtt{\mathbf{g}}$ satisfies Einstein equations. This is done in Section 3.
Moreover, it is shown in Section 7 that the formalism can also be interpreted
in terms of a $\mathtt{\mathbf{g}}$-compatible teleparallel connection in $M$,
which produces in a an clear and elegant way the so called teleparallel
equivalent of General Relativity. The situation here is somewhat analogous to
the one in the following example. Suppose you have a punctured sphere
$\mathring{S}^{2}$ that lives in $\mathbb{R}^{3}$. Which is the best geometry
that you can use in $\mathring{S}^{2}$? Well, the answer depends on the
applications you have in mind. It may be useful for some problems (computation
of \ curves of minimum length (geodesics)) to use a Riemannian geometrical
structure $(\mathring{S}^{2},g,\overset{LC}{D})$, where $g$ is pullback on
$\mathring{S}^{2}$ of the Euclidean metric on $\mathbb{R}^{3}$, and
$\overset{LC}{D}$ is the Levi-Civita connection of $g$, or it may be more
useful (e.g., for sailors) to use the structure $(\mathring{S}^{2}%
,g,\overset{N}{D})$ where $\overset{N}{D}$ is the Nunes connection (also
called navigator, or Columbus connection \cite{rodoliv2006}). There are still
some problems \cite{oneill} where the use of a Euclidean geometry on
$\mathring{S}^{2}$ is the most useful one. This last geometry defines the so
called stereographic sphere $(\mathring{S}^{2},g^{\prime},\overset
{LC}{D^{\prime}})$ . In it a metric $g^{\prime}$ is defined by pullback of the
Euclidean metric of a tangent plane at the south pole with the diffeomorphism
map defined by stereographic projection map (from the north pole). The
connection $\overset{LC}{D^{\prime}}$ is defined as the Levi-Civita connection
of $g^{\prime}$.

To present the details of our theory we start by introducing $\{%
\slx
^{\mathbf{\mu}}\}$, which are global coordinate functions\footnote{The
coordinates of $\mathfrak{e}\in M$ in Einstein-Lorentz-Poincar\'{e} gauge are
$\{x^{\mu}\}:=\{%
\slx
^{\mu}(\mathfrak{e})\}$.} in Einstein-Lorentz-Poincar\'{e} gauge for $M$
associated to an arbitrary inertial reference frame\footnote{An inertial
reference frame satisfies $DI=0$. See \cite{rodoliv2006} for details.}
$I=\partial/\partial x^{0}\in\sec TM$. Let $\{e_{\mathbf{a}}=\delta
_{\mathbf{a}}^{\mu}\partial/\partial x^{\mu}\}$ be an orthonormal basis for
$TM$ and $\{\vartheta^{\mathbf{a}}\}$ the corresponding dual basis for
$T^{\ast}M$. We have $\vartheta^{\mathbf{a}}=\delta_{\mu}^{\mathbf{a}}dx^{\mu
}$, $\mathbf{a}=0,1,2,3$, and we take\footnote{$\bigwedge\nolimits^{p}T^{\ast
}M$ denotes the bundle of $p$-forms, $\bigwedge T^{\ast}M=%
{\displaystyle\bigoplus\nolimits_{p=0}^{4}}
\bigwedge\nolimits^{p}T^{\ast}M$ is the bundle of multiform fields,
$\mathcal{C}\ell(M,\mathtt{\eta})$ denotes the Clifford bundle of differential
forms. The symbol $\sec$ means section. All \ `tricks of the trade' necessary
for performing the calculations of the present paper are described in
\cite{rodoliv2006}.} $\vartheta^{\mathbf{a}}\in\sec\bigwedge\nolimits^{1}%
T^{\ast}M\hookrightarrow\sec\mathcal{C}\ell(M,\mathtt{\eta})$. Of course, we
have
\begin{equation}%
\mbox{\boldmath{$\eta$}}%
=\eta_{\mathbf{ab}}\vartheta^{\mathbf{a}}\otimes\vartheta^{\mathbf{b}%
}.\label{zero}%
\end{equation}

\textbf{Assumption 1}: A non trivial gravitational field is represented by a
basis $\{\mathfrak{g}^{\mathbf{a}}\}$ of $T^{\ast}M$, defining a. $\eta
$-orthonormal coframe bundle for $M$, which is \textit{not} a coordinate
coframe in all $M$ and such that\footnote{Take notice that $%
\mbox{\boldmath{$\eta$}}%
=\eta_{\mathbf{ab}}\vartheta^{\mathbf{a}}\otimes\vartheta^{\mathbf{b}}\neq
\eta$.=$\eta_{\mathbf{ab}}\mathfrak{g}^{\mathbf{a}}\otimes\mathfrak{g}%
^{\mathbf{b}}$} $\eta=\eta_{\mathbf{ab}}\mathfrak{g}^{\mathbf{a}}%
\otimes\mathfrak{g}^{\mathbf{b}}.$In what follows we suppose moreover that
$\mathfrak{g}^{\mathbf{a}}\in\sec\bigwedge\nolimits^{1}T^{\ast}%
M\hookrightarrow\sec\mathcal{C}\ell(M,\mathtt{\eta})$. The fields
$\mathfrak{g}^{\mathbf{a}}$ in a region of $M$ generated by the matter fields
with Lagrangian density $\mathcal{L}_{m}^{M}$ are described by a Lagrangian
density%
\begin{equation}
\mathcal{L=L}_{g}^{M}+\mathcal{L}_{m}^{M},\label{8.5neww}%
\end{equation}
where%
\begin{equation}
\mathcal{L}_{g}^{M}=-\frac{1}{2}d\mathfrak{g}^{\mathbf{a}}\wedge
\underset{\mathbf{\eta}}{\star}d\mathfrak{g}_{\mathbf{a}}+\frac{1}{2}%
\underset{\mathbf{\eta}}{\delta}\mathfrak{g}^{\mathbf{a}}\wedge\underset
{\mathbf{\eta}}{\star}\underset{\mathbf{\eta}}{\delta}\mathfrak{g}%
_{\mathbf{a}}+\frac{1}{4}d\mathfrak{g}^{\mathbf{a}}\wedge\mathfrak{g}%
_{\mathbf{a}}\wedge\underset{\mathbf{\eta}}{\star}d\mathfrak{g}^{\mathbf{b}%
}\wedge\mathfrak{g}_{\mathbf{b}},\label{10hyp2}%
\end{equation}
is invariant under local Lorentz transformations\footnote{We observe that
various coefficients in Eq.(\ref{10hyp2}) have been selected in order for
$\mathcal{L}_{g}^{M}$ to be invariant under arbitrary local Lorentz
transformations. This means, as the reader may verifiy that under the
transformation \ $\mathfrak{g}^{\mathbf{a}}\mapsto u\mathfrak{g}^{\mathbf{a}%
}u^{-1}$, $u\in\sec\mathrm{Spin}_{1,3}^{e}(M,\mathtt{\eta})\hookrightarrow
\sec\mathcal{C}\ell(M,\mathtt{\eta})$, $\mathcal{L}_{g}^{M}$ is invariant
modulo an exact form.}, which is a kind of gauge freedom, a crucial ingredient
of our theory, as will be clear in a while. Moreover, $\underset{\mathbf{\eta
}}{\star}$ refers to the Hodge dual defined by $\eta=\eta_{\mathbf{ab}%
}\mathfrak{g}^{\mathbf{a}}\otimes\mathfrak{g}^{\mathbf{b}}$.

The $\mathfrak{g}^{\mathbf{a}}$ couple universally to the matter fields in
such a way that the energy momentum $1$-form of the matter fields are given by%
\begin{equation}
-\underset{\mathbf{\eta}}{\star}T_{\mathbf{a}}^{M}=\underset{\mathbf{\eta}%
}{\star}T_{\mathbf{a}}^{M}=\frac{\partial\mathcal{L}_{m}^{M}}{\partial
\mathfrak{g}^{\mathbf{a}}}. \label{10hyp2'}%
\end{equation}
\medskip

We see that each one of the fields $\mathfrak{g}^{\mathbf{a}}$ \ in
Eq.(\ref{10hyp2}) resembles a potential of an electromagnetic field. Indeed,
the first term is of the Yang-Mills type, the second term is a kind of gauging
fixing term, for indeed, $\underset{\mathbf{\eta}}{\delta}\mathfrak{g}%
^{\mathbf{a}}=0$ is analogous to the \textit{Lorenz} condition for the gauge
potential of the electromagnetic potential and finally the third term is a
self-interacting term, which is proportional to the square of the
total\ `vorticity' $\Omega=d\mathfrak{g}^{\mathbf{a}}\wedge\mathfrak{g}%
_{\mathbf{a}}$ associated to the $1$-form fields $\mathfrak{g}^{\mathbf{a}}$ .
We will derive in Section 4 Maxwell like equations for the gravitational
fields. Comparison of our equations with the ones found by
\cite{itina,itin,kaniel} are mentioned

We see that in our formulation of the theory of gravitational field there is
until now no mention to a Lorentzian spacetime structure $(M\simeq
\mathbb{R}^{4},\mathtt{\mathbf{g}},\nabla,\tau_{\mathtt{\mathbf{g}}}%
,\uparrow)$. Such structure enters the game by supposing that the most general
deformation of the Lorentz vacuum can be described by a diffeomorphism
$\mathtt{h}:M\rightarrow M$, $\mathfrak{e}\mapsto\mathtt{h}\mathfrak{e}$, and
a related gauge metric extensor field $h$, which are introduced next.

\section{$\mathcal{C\ell}(M,\mathtt{\eta})$, $\mathcal{C\ell}(M,\mathtt{g})$,
\texttt{h} and $h$}

\subsection{Enter \texttt{h}}

\textbf{Assumption 2}: Every physically acceptable gravitational
$\{\mathfrak{g}^{\mathbf{a}}\}$ induces a metric field $\mathtt{\mathbf{g}}%
\in\sec T_{2}^{0}M$ which is a Cauchy-Green like tensor \cite{frankel}, i.e.,
it is the pullback of the metric\footnote{Take notice that $%
\mbox{\boldmath{$\eta$}}%
=\eta_{\mathbf{ab}}\vartheta^{\mathbf{a}}\otimes\vartheta^{\mathbf{b}}\neq
\eta$.$=\eta_{\mathbf{ab}}\mathfrak{g}^{\mathbf{a}}\otimes\mathfrak{g}%
^{\mathbf{b}}$} $\eta=\eta_{\mathbf{ab}}\mathfrak{g}^{\mathbf{a}}%
\otimes\mathfrak{g}^{\mathbf{b}}$ under a diffeomorphism $\mathtt{h}%
:M\rightarrow M$, $\mathfrak{e}\mapsto\mathtt{h}\mathfrak{e}$. We have,%
\begin{align}
\mathtt{\mathbf{g}} &  =\mathtt{h}^{\ast}\eta=\eta\mathbf{_{\mathbf{ab}}%
}\theta^{\mathbf{a}}\otimes\theta^{\mathbf{b}}\mathbf{,}\label{10hyp}\\
\mathfrak{g}^{\mathbf{a}} &  =\mathtt{h}^{\ast-1}\theta^{\mathbf{a}%
}.\label{10hypa}%
\end{align}
\medskip

To show that our assumptions imply indeed that $\mathtt{\mathbf{g}}$ satisfies
Einstein equations as stated above, we need to prove that $\mathcal{L}_{g}%
^{M}$ is equivalent to the Einstein-Hilbert Lagrangian. This will be done
after we prove Proposition \label{Lg}, which needs some preliminaries.

\subsection{Enter $h$}

Consider the Clifford bundles of nonhomogeneous multiform fields
$\mathcal{C\ell}(M,\mathtt{\eta})$ and $\mathcal{C\ell}(M,\mathtt{g})$. In
$\mathcal{C\ell}(M,\mathtt{\eta})$, where $\mathtt{\eta}$ refers to the
standard metric on the cotangent bundle associated to $%
\mbox{\boldmath{$\eta$}}%
=\eta_{\mathbf{ab}}\mathfrak{\vartheta}^{\mathbf{a}}\otimes\mathfrak{\vartheta
}^{\mathbf{b}}$, the Clifford product will be denoted by juxtaposition of
symbols, the scalar product by $\cdot$ and the contractions by $\lrcorner$ and
$\llcorner$ and by $\star$ we denote the Hodge dual. The Clifford product in
$\mathcal{C\ell}(M,\mathtt{g})$ will be denoted by the symbol $\vee$, the
scalar product will be denoted by $\bullet$ and the contractions by
$\underset{\mathtt{\mathbf{g}}}{\lrcorner}$ and $\underset{\mathtt{\mathbf{g}%
}}{\llcorner}$ while by $\underset{\mathtt{\mathbf{g}}}{\star}$ we denote the
Hodge dual operator associated to \texttt{$\mathbf{g}$}.

Let $\{\mathbf{e}_{a}\}$ be a non coordinate basis of $TM$ dual to the cobasis
$\{\theta^{\mathbf{a}}\}$. We take the $\theta^{\mathbf{a}}$ as sections of
the Clifford bundle $\mathcal{C}\ell(M,\mathtt{\eta})$, i.e., $\theta
^{\mathbf{a}}\in\sec\bigwedge\nolimits^{1}T^{\ast}M\hookrightarrow
\sec\mathcal{C}\ell(M,\mathtt{\eta})$. In this basis where according to
Eq.(\ref{10hyp}) \texttt{$\mathbf{g}$ }$=\eta_{\mathbf{ab}}\theta^{\mathbf{a}%
}\otimes\theta^{\mathbf{b}}$ we have that
\begin{equation}
\eta=g_{\mathbf{ab}}\theta^{\mathbf{a}}\otimes\theta^{\mathbf{b}},
\label{8.2bis}%
\end{equation}
and moreover $\mathtt{g}$\texttt{ }$\in\sec T_{0}^{2}M$ is given by
\begin{equation}
\mathtt{g}=\eta^{\mathbf{ab}}\mathbf{e}_{\mathbf{a}}\otimes\mathbf{e}%
_{\mathbf{b}}. \label{8.1a}%
\end{equation}
The cobasis $\{\vartheta^{\mathbf{a}}\}$ defines a Clifford product in
$\mathcal{C}\ell(M,\mathtt{\eta})$ by
\begin{equation}
\vartheta^{\mathbf{a}}\vartheta^{\mathbf{b}}+\vartheta^{\mathbf{b}}%
\vartheta^{\mathbf{a}}=2\eta^{\mathbf{ab}}, \label{8.3}%
\end{equation}
and taking into account that the cobasis $\{\theta^{\mathbf{a}}\}$ defines a
\textit{deformed} Clifford product $\vee$ in $\mathcal{C}\ell(M,\mathtt{\eta
})$ (see details in \cite{rodoliv2006,mofero}) generating a representation of
the Clifford bundle $\mathcal{C\ell}(M,\mathtt{g})$ we can write
\begin{align}
\theta^{\mathbf{a}}\vee\theta^{\mathbf{b}}  &  =\theta^{\mathbf{a}}%
\bullet\theta^{\mathbf{b}}+\theta^{\mathbf{a}}\wedge\theta^{\mathbf{b}%
},\nonumber\\
\theta^{\mathbf{a}}\vee\theta^{\mathbf{b}}+\theta^{\mathbf{b}}\vee
\theta^{\mathbf{a}}  &  =2\eta^{\mathbf{ab}}. \label{8.4}%
\end{align}
Then, as proved, e.g., in \cite{rodoliv2006,mofero} there exist $(1,1)$%
-extensor fields $g$ and $h$ such that
\begin{equation}
\mathtt{g}(\theta^{\mathbf{a}},\theta^{\mathbf{b}})=\theta^{\mathbf{a}}%
\bullet\theta^{\mathbf{b}}=\theta^{\mathbf{a}}\cdot g(\theta^{\mathbf{b}%
})=h(\theta^{\mathbf{a}})\cdot h(\theta^{\mathbf{b}})=\eta^{\mathbf{ab}}.
\label{8.5}%
\end{equation}

The gauge metric extensor $h:\sec%
{\displaystyle\bigwedge\nolimits^{1}}
T^{\ast}M\rightarrow\sec%
{\displaystyle\bigwedge\nolimits^{1}}
T^{\ast}M$ is defined by
\begin{equation}
h(\theta^{\mathbf{a}})=\vartheta^{\mathbf{a}}. \label{8.5new}%
\end{equation}

\subsection{Relation Between $h$ and $\mathtt{h}^{\ast}$}

\textbf{ }Recall that \textbf{Assumption 2} says that every physically
acceptable $\mathtt{\mathbf{g}}$ is a Cauchy-Green like tensor \cite{frankel},
i.e., it is the pullback of the metric $\eta=\eta_{\mathbf{ab}}\mathfrak{g}%
^{\mathbf{a}}\otimes\mathfrak{g}^{\mathbf{b}}$ under a diffeomorphism
$\mathtt{h}:M\rightarrow M$, $\mathfrak{e}\mapsto\mathtt{h}\mathfrak{e}$.

Introduce Riemann normal coordinates functions $\{%
\sly
^{\mu}\}$ for $M$ such that%
\begin{equation}%
\sly
^{\mu}(\mathtt{h}\mathfrak{e)}=y^{\mu}, \label{8.27'}%
\end{equation}
and being $%
\slx
^{\mu}$ the coordinates in the the Einstein-Lorentz-Poincar\'{e} gauge for $M$
already introduced and obeying $%
\slx
^{\mu}(\mathfrak{e})=x^{\mu}$, we have
\begin{equation}
\left.  \eta\right\vert _{\mathtt{h}\mathfrak{e}}\mathbf{=}\eta_{\mathbf{ab}%
}\delta_{\mu}^{\mathbf{a}}\delta_{\nu}^{\mathbf{b}}dy^{\mathbf{\mu}}\otimes
dy^{\nu}\text{, }\left.  \mathtt{\mathbf{g}}\right\vert _{\mathfrak{e}}%
=g_{\mu\nu}dx^{\mu}\otimes dx^{\nu}. \label{8.27a}%
\end{equation}
Moreover, let $y^{\mu}=\mathtt{h}^{\mu}(x^{\nu})$ be the coordinate
expression\footnote{Recall that the $\mathtt{h}^{\mu}$ are assumed invertible
differentiable functions.} for \texttt{h.} Since%
\[
\left.  \mathtt{h}^{\ast}\eta\mathbf{(}\delta_{\mathbf{a}}^{\mu}%
\mathbf{\partial/\partial}x^{\mu},\delta_{\mathbf{b}}^{\nu}\mathbf{\partial
/\partial}x^{\nu}\mathbf{)}\right\vert _{\mathfrak{e}}\mathbf{=}\left.
\eta\mathbf{(}\delta_{\mathbf{a}}^{\mu}\mathtt{h}_{\ast}\mathbf{\partial
/\partial}x^{\mu},\delta_{\mathbf{b}}^{\nu}\mathtt{h}_{\ast}\mathbf{\partial
/\partial}x^{\nu}\mathbf{)}\right\vert _{\mathtt{h}\mathfrak{e}},
\]%
\begin{equation}
\mathtt{\mathbf{g}}=\mathtt{h}^{\ast}\eta=\eta_{\mathbf{ab}}\delta_{\alpha
}^{\mathbf{a}}\delta_{\beta}^{\mathbf{b}}\frac{\partial y^{\alpha}}{\partial
x^{\mu}}\frac{\partial y^{\beta}}{\partial x^{\nu}}dx^{\mu}\otimes dx^{\nu}
\label{8.27b}%
\end{equation}
with
\begin{equation}
g_{\mu\nu}=\eta_{\mathbf{ab}}\delta_{\alpha}^{\mathbf{a}}\delta_{\beta
}^{\mathbf{b}}\frac{\partial y^{\alpha}}{\partial x^{\mu}}\frac{\partial
y^{\beta}}{\partial x^{\nu}}. \label{8.27c}%
\end{equation}

Now, take notice that at $\mathfrak{e}$,$\ \{\mathbf{f}_{\mathbf{a}}\}$,
$\mathbf{f}_{\mathbf{a}}=\delta_{\mathbf{a}}^{\mu}\mathtt{h}_{\ast}%
^{-1}\mathbf{\partial/\partial}y^{\mu}=\delta_{\mathbf{a}}^{\mu}\frac{\partial
x^{\nu}}{\partial y^{\mu}}\frac{\partial}{\partial x^{\nu}}$ is not (in
general) a coordinate basis for $TM$. It is also not $\eta$%
-orthonormal\footnote{Indeed, $\eta(\mathbf{e}_{\mathbf{a}},\mathbf{e}%
_{\mathbf{b}})=\delta_{\mathbf{a}}^{\mu}\delta_{\mathbf{b}}^{\nu}%
\frac{\partial y^{\alpha}}{\partial x^{\mu}}\frac{\partial y^{\beta}}{\partial
x^{\nu}}\eta_{\alpha\beta}$.}. The dual basis of $\{\mathbf{f}_{\mathbf{a}}\}$
at $\mathfrak{e}$ is $\{\left.  \sigma^{\mathbf{a}}\right\vert _{\mathfrak{e}%
}\}$\textbf{, }with\textbf{ }$\left.  \sigma^{\mathbf{a}}\right\vert
_{\mathfrak{e}}=\left.  \delta_{\mu}^{\mathbf{a}}\frac{\partial y^{\mu}%
}{\partial x^{\nu}}dx^{\nu}\right\vert _{\mathfrak{e}}=\mathtt{h}^{\ast
}(\left.  \delta_{\mu}^{\mathbf{a}}dy^{\mu}\right\vert _{\mathtt{h}%
\mathfrak{e}})$. Then it exists an extensor field $\check{h}\ $differing from
$h$ by a Lorentz extensor, i.e., $\check{h}=h\Lambda$ such that$\ \left.
\sigma^{\mathbf{a}}\right\vert _{\mathfrak{e}}=\left.  \check{h}^{-1}%
(\delta_{\mu}^{\mathbf{a}}dy^{\mu})\right\vert _{\mathfrak{e}}=\left.
\check{h}_{\mu}^{-1\mathbf{a}}dy^{\mu}\right\vert _{\mathfrak{e}}$, we have
for any $\mathfrak{e}\in M$, $\ $%
\begin{equation}
\delta_{\alpha}^{\mathbf{a}}\frac{\partial y^{\alpha}}{\partial x^{\mu}%
}=\check{h}_{\mu}^{-1\mathbf{a}}.\label{8.27dd}%
\end{equation}

To determine $\check{h}$ we proceed as follows. Suppose $\mathtt{\mathbf{g}%
}=\eta_{\mathbf{ab}}\sigma^{\mathbf{a}}\otimes\sigma^{\mathbf{b}}$ is known.
Let ($v_{i},\lambda_{i})$ be respectively the eigen-covectors and the
eigenvalues of $g$, i.e., $g(v_{i})=\lambda_{i}v_{i}$ (no sum in $i$) and
$\{\vartheta^{\mathbf{a}}\}$ the $\eta$-orthonormal coordinate basis for
$T^{\ast}M$ introduced above. Then, since $g=\check{h}^{\dagger}\check{h}$ we
immediately have
\begin{equation}
\check{h}(v_{i})=\sqrt{\left\vert \lambda_{i}\right\vert }\mathtt{\eta}%
(v_{i},\vartheta_{\mathbf{a}})\vartheta^{\mathbf{a}}, \label{8.29bis}%
\end{equation}
which then determines the extensor field $h$ (modulus a \ local Lorentz
rotation) at any spacetime point, and thus the diffeomorphism \texttt{h}
(modulus a local Lorentz rotation).

\section{Enter $\nabla$}

Now, any other $\mathtt{\mathbf{g}}$-orthonormal \ non coordinate cobasis
$\{\theta^{\mathbf{a}}\}$ is related by a Lorentz extensor field $\Lambda$,
(i.e., $\Lambda^{\dagger}g\Lambda=g$) to the cobasis $\{\sigma^{\mathbf{a}}\}$
by $\theta^{\mathbf{a}}=\Lambda(\sigma^{\mathbf{a}})=U\sigma^{\mathbf{a}%
}U^{-1}$, $U\in\sec\mathrm{Spin}_{1,3}^{e}(M,\mathtt{g})\hookrightarrow
\sec\mathcal{C}\ell(M,\mathtt{g})$. Then, $\mathtt{\mathbf{g}}=\eta
_{\mathbf{ab}}\sigma^{\mathbf{a}}\otimes\sigma^{\mathbf{b}}=$ $\eta
_{\mathbf{ab}}\theta^{\mathbf{a}}\otimes\theta^{\mathbf{b}}$ and
\textbf{assumption 2} says that\textbf{ }$\mathtt{\mathbf{g}}=\eta
_{\mathbf{ab}}\theta^{\mathbf{a}}\otimes\theta^{\mathbf{b}}=\mathtt{h}^{\ast
}\eta\mathbf{=}\mathtt{h}^{\ast}(\eta_{\mathbf{ab}}\mathfrak{g}^{\mathbf{a}%
}\otimes\mathfrak{g}^{\mathbf{b}})$, i.e., $\mathfrak{g}^{\mathbf{a}%
}=\mathtt{h}^{\ast-1}\theta^{\mathbf{a}}$. Taking into account also that
$\mathfrak{g}_{\mathbf{a}}=\eta_{\mathbf{ab}}\mathfrak{g}^{\mathbf{b}}$, we
can write using the identities%

\begin{align}
d\mathtt{h}^{\ast}K  &  =\mathtt{h}^{\ast}dK\nonumber\\
\underset{\mathtt{\mathbf{g}}}{\star}d\underset{\mathtt{\mathbf{g}}}{\star
}\mathtt{h}^{\ast}K  &  =\text{ }\mathtt{h}^{\ast}\underset{\eta}{\star
}d\underset{\eta}{\star}K,\nonumber\\
\mathtt{h}^{\ast}L\underset{\mathtt{\mathbf{g}}}{\lrcorner}\mathtt{h}^{\ast}K
&  =\mathtt{h}^{\ast}(L\underset{\eta}{\lrcorner}K), \label{ndif4}%
\end{align}
valid for any $L\in\sec%
{\displaystyle\bigwedge\nolimits^{r}}
T^{\ast}M$, $K\in\sec%
{\displaystyle\bigwedge\nolimits^{p}}
T^{\ast}M$ , $r\leq p$, that
\begin{align}
d\theta^{\mathbf{a}}\wedge\underset{\mathtt{\mathbf{g}}}{\star}d\theta
_{\mathbf{a}}  &  =\mathtt{h}^{\ast}d\mathtt{h}^{\ast-1}\theta^{\mathbf{a}%
}\wedge\underset{\mathtt{\mathbf{g}}}{\star}d\theta_{\mathbf{a}}\nonumber\\
&  =\mathtt{h}^{\ast}d\mathtt{h}^{\ast-1}\theta^{\mathbf{a}}\wedge
\mathtt{h}^{\ast}\underset{\mathbf{\eta}}{\star}d\mathtt{h}^{\ast-1}%
\theta_{\mathbf{a}}\nonumber\\
&  \mathtt{h}^{\ast}(d\mathtt{h}^{\ast-1}\theta^{\mathbf{a}}\wedge
\underset{\eta}{\star}d\mathtt{h}^{\ast-1}\theta_{\mathbf{a}})\nonumber\\
&  =\mathtt{h}^{\ast}(d\mathfrak{g}^{\mathbf{a}}\wedge\underset{\eta}{\star
}d\mathfrak{g}_{\mathbf{a}}) \label{8.27e}%
\end{align}

and \
\begin{align}
\underset{\mathbf{\eta}}{\delta}\theta^{\mathbf{a}}\wedge\underset
{\mathbf{\eta}}{\star}\underset{\mathbf{\eta}}{\delta}\theta_{\mathbf{a}}  &
=\mathtt{h}^{\ast}(\underset{\eta}{\delta}\mathfrak{g}^{\mathbf{a}}%
\wedge\underset{\mathbf{\eta}}{\star}\underset{\mathbf{\eta}}{\delta
}\mathfrak{g}_{\mathbf{a}}),\nonumber\\
d\theta^{\mathbf{a}}\wedge\theta_{\mathbf{a}}\wedge\underset
{\mathtt{\mathbf{g}}}{\star}d\theta^{\mathbf{b}}\wedge\theta_{\mathbf{b}}  &
=\mathtt{h}^{\ast}(d\mathfrak{g}^{\mathbf{a}}\wedge\mathfrak{g}_{\mathbf{a}%
}\wedge\underset{\mathbf{\eta}}{\star}d\mathfrak{g}^{\mathbf{b}}%
\wedge\mathfrak{g}_{\mathbf{b}}).
\end{align}

Then, the Lagragian for the gravitational field becomes
\begin{equation}
\mathcal{L}_{g}^{M}=\mathtt{h}^{\ast-1}\mathcal{L}_{g}, \label{8.1biss}%
\end{equation}
\ where%
\begin{equation}
\mathcal{L}_{g}=-\frac{1}{2}d\theta^{\mathbf{a}}\wedge\underset
{\mathtt{\mathbf{g}}}{\star}d\theta_{\mathbf{a}}+\frac{1}{2}\underset
{\mathtt{\mathbf{g}}}{\delta}\theta^{\mathbf{a}}\wedge\underset
{\mathtt{\mathbf{g}}}{\star}\underset{\mathtt{\mathbf{g}}}{\delta}%
\theta_{\mathbf{a}}+\frac{1}{4}\left(  d\theta^{\mathbf{a}}\wedge
\theta_{\mathbf{a}}\right)  \wedge\underset{\mathtt{\mathbf{g}}}{\star}\left(
d\theta^{\mathbf{b}}\wedge\theta_{\mathbf{b}}\right)  . \label{8.1bis}%
\end{equation}
We now show that $\mathcal{L}_{g}$ differs from the Einstein-Hilbert
Lagrangian by an exact differential\footnote{Observe that $-\frac{1}%
{2}(d\theta^{\mathbf{a}}\wedge\underset{\mathtt{\mathbf{g}}}{\star}%
d\theta_{\mathbf{a}}-\underset{\mathtt{\mathbf{g}}}{\delta}\theta^{\mathbf{a}%
}\wedge\underset{\mathtt{\mathbf{g}}}{\star}\underset{\mathtt{\mathbf{g}}%
}{\delta}\theta_{\mathbf{a}})=\frac{1}{4}\left(  d\theta^{\mathbf{a}}%
\wedge\theta_{\mathbf{b}}\right)  \wedge\underset{\mathtt{\mathbf{g}}}{\star
}\left(  d\theta^{\mathbf{b}}\wedge\theta_{\mathbf{a}}\right)  $ and
$-\frac{1}{2}\left(  d\theta^{\mathbf{a}}\wedge\theta_{\mathbf{b}}\right)
\wedge\underset{\mathtt{\mathbf{g}}}{\star}\left(  d\theta^{\mathbf{b}}%
\wedge\theta_{\mathbf{a}}\right)  +\frac{1}{4}\left(  d\theta^{\mathbf{a}%
}\wedge\theta_{\mathbf{a}}\right)  \wedge\underset{\mathtt{\mathbf{g}}}{\star
}\left(  d\theta^{\mathbf{b}}\wedge\theta_{\mathbf{b}}\right)  $ is known to
differ from the Einstein-Hilbert Lagrangian by an exact
differential\cite{thirring,roldaorod}}. It is at this point that the
Levi-Civita connection $\nabla$ of $\mathtt{\mathbf{g}}$ comes to play.
Indeed, we introduce the connections $1$-forms $\omega_{\mathbf{b}%
}^{\mathbf{a}}$ and the curvature $2$-forms in the cobasis $\{\theta
^{\mathbf{a}}\}$ of the Lorentzian spacetime structure\ $(M\simeq
\mathbb{R}^{4},\mathtt{\mathbf{g}},\nabla,\tau_{\mathtt{\mathbf{g}}}%
,\uparrow)$ through Cartan's structure equations%
\begin{equation}
d\theta^{\mathbf{a}}+\omega_{\mathbf{b}}^{\mathbf{a}}\wedge\theta^{\mathbf{b}%
}=0, \label{cartan1}%
\end{equation}%
\begin{equation}
\mathcal{R}_{\mathbf{b}}^{\mathbf{a}}=d\omega_{\mathbf{b}}^{\mathbf{a}}%
+\omega_{\mathbf{c}}^{\mathbf{a}}\wedge\omega_{\mathbf{b}}^{\mathbf{c}}.
\label{cartan2}%
\end{equation}

Next, we suppose that all objects in Eq.(\ref{cartan1}) and Eq.(\ref{cartan2})
are forms in $\mathcal{C}\ell(M,\mathtt{g})$ represented as explained above in
$\mathcal{C}\ell(M,\mathtt{\eta})$. Under this condition Eq.(\ref{cartan1})
can be easily inverted, i.e., we get%
\begin{equation}
\omega^{\mathbf{cd}}=\frac{1}{2}\left[  \theta^{\mathbf{d}}\underset
{\mathtt{\mathbf{g}}}{\lrcorner}d\theta^{\mathbf{c}}-\theta^{\mathbf{c}%
}\underset{\mathtt{\mathbf{g}}}{\lrcorner}d\theta^{\mathbf{d}}+\theta
^{\mathbf{c}}\underset{\mathtt{\mathbf{g}}}{\lrcorner}\left(  \theta
^{\mathbf{d}}\underset{\mathtt{\mathbf{g}}}{\lrcorner}d\theta_{\mathbf{a}%
}\right)  \theta^{\mathbf{a}}\right]  . \label{8.19}%
\end{equation}

\begin{remark}
It is crucial to observe that the Levi-Civita connection $\nabla$ of
$\mathtt{\mathbf{g}}$ such that $\nabla_{\mathbf{e}_{\mathbf{a}}}%
\theta^{\mathbf{b}}\mathbf{=-}L_{\mathbf{ac}}^{\mathbf{b}}\theta^{\mathbf{c}}%
$, $\omega_{\mathbf{a}}^{\mathbf{b}}=-L_{\mathbf{ac}}^{\mathbf{b}}%
\theta^{\mathbf{c}}$ is not \emph{(}of course\emph{)} the pullback of the
Levi-Civita connection $D$ of $%
\mbox{\boldmath{$\eta$}}%
$ where $D_{\mathbf{e}_{\mathbf{a}}}\theta^{\mathbf{b}}=-\Gamma_{\mathbf{ac}%
}^{\mathbf{b}}\theta^{\mathbf{c}}$. Indeed, as the reader may easily verify,
if that was the case the Riemann tensor of $\nabla$ would be null. Recall that
since we supposed that $d\mathfrak{g}^{\mathbf{a}}=-\frac{1}{2}c_{\mathbf{kb}%
}^{\prime\mathbf{a}}\mathfrak{g}^{\mathbf{k}}\wedge\mathfrak{g}^{\mathbf{b}%
}\neq0$, we can introduce on $M$ connection 1-forms $\omega_{\mathbf{b}%
}^{\prime\mathbf{a}}:=\frac{1}{2}c_{\mathbf{kb}}^{\prime\mathbf{a}%
}\mathfrak{g}^{\mathbf{k}}\ $such that they define a connection $\nabla
^{\prime}$wich is metric compatible with $\eta=\eta_{\mathbf{ab}}%
\mathfrak{g}^{\mathbf{a}}\otimes\mathfrak{g}^{\mathbf{b}}$, \emph{(}%
$\nabla_{\mathfrak{e}_{\mathbf{a}}}^{\prime}\eta=0\emph{)}$, for which the
torsion tensor is $\Theta^{\prime\mathbf{a}}=d\mathfrak{g}^{\mathbf{a}}%
+\omega_{\mathbf{b}}^{\prime\mathbf{a}}\wedge\mathfrak{g}^{\mathbf{b}}=0$ and
$\mathcal{R}_{\mathbf{b}}^{\prime\mathbf{a}}=d\omega_{\mathbf{b}}%
^{\prime\mathbf{a}}+\omega_{\mathbf{c}}^{\prime\mathbf{a}}\wedge
\omega_{\mathbf{b}}^{\prime\mathbf{c}}\neq0$. Now, $\nabla$ can be wieved as
the pullback of $\ \nabla^{\prime}$ and of course, we have that the connection
1-forms of $\ \nabla$ are $\omega_{\mathbf{b}}^{\mathbf{a}}=\mathtt{h}^{\ast
}\omega_{\mathbf{b}}^{\prime\mathbf{a}}=\frac{1}{2}c_{\mathbf{kb}}%
^{\prime\mathbf{a}}\mathfrak{\theta}^{\mathbf{k}}$ and $\nabla
\mathtt{\mathbf{g}}=0$. Moreover, $\Theta^{\mathbf{a}}=\mathtt{h}^{\ast}%
\Theta^{\prime\mathbf{a}}=0$, but $\mathcal{R}_{\mathbf{b}}^{\mathbf{a}%
}=\mathtt{h}^{\ast}\mathcal{R}_{\mathbf{b}}^{\prime\mathbf{a}}\neq0$.\ 
\end{remark}

\subsection{Relation Between $\mathcal{L}_{EH}$ and $\mathcal{L}_{g}$}

Recall that the classical Einstein-Hilbert Lagrangian in appropriate
(geometrical) units is
\begin{equation}
\mathcal{L}_{EH}=\frac{1}{2}R\tau_{\mathtt{\mathbf{g}}}, \label{8.6}%
\end{equation}
where $R=\eta^{\mathbf{cd}}R_{\mathbf{cd}}$ is the scalar curvature. Now,
observe that we can write $\mathcal{L}_{EH}$ as
\begin{equation}
\mathcal{L}_{EH}=\frac{1}{2}\mathcal{R}_{\mathbf{cd}}\wedge\underset
{\mathtt{\mathbf{g}}}{\star}(\theta^{\mathbf{c}}\wedge\theta^{\mathbf{d}}),
\label{8.7}%
\end{equation}
where
\begin{equation}
\mathcal{R}_{\mathbf{d}}^{\mathbf{a}}=d\omega_{\mathbf{b}}^{\mathbf{a}}%
+\omega_{\mathbf{c}}^{\mathbf{a}}\wedge\omega_{\mathbf{b}}^{\mathbf{c}},
\label{8.7a}%
\end{equation}
are the curvature $2$-forms. Indeed, using well known identities (see, e.g.,
Chapter 2 of \cite{rodoliv2006}), we have
\begin{align}
\mathcal{R}_{\mathbf{cd}}\wedge\underset{\mathtt{\mathbf{g}}}{\star}%
(\theta^{\mathbf{c}}\wedge\theta^{\mathbf{d}})  &  =(\theta^{\mathbf{c}}%
\wedge\theta^{\mathbf{d}})\wedge\underset{\mathtt{\mathbf{g}}}{\star
}\mathcal{R}_{\mathbf{cd}}=-\theta^{\mathbf{c}}\wedge\underset
{\mathtt{\mathbf{g}}}{\star}(\theta^{\mathbf{d}}\underset{\mathtt{\mathbf{g}}%
}{\lrcorner}\mathcal{R}_{\mathbf{cd}})\nonumber\\
&  =-\underset{\mathtt{\mathbf{g}}}{\star}[\theta^{\mathbf{c}}\underset
{\mathtt{\mathbf{g}}}{\lrcorner}(\theta^{\mathbf{d}}\underset
{\mathtt{\mathbf{g}}}{\lrcorner}\mathcal{R}_{\mathbf{cd}})], \label{8.8}%
\end{align}
and since%
\begin{align}
\theta^{\mathbf{d}}\underset{\mathtt{\mathbf{g}}}{\lrcorner}\mathcal{R}%
_{\mathbf{cd}}  &  =\frac{1}{2}R_{\mathbf{cdab}}\theta^{\mathbf{d}}%
\underset{\mathtt{\mathbf{g}}}{\lrcorner}(\theta^{\mathbf{a}}\wedge
\theta^{\mathbf{b}})=\frac{1}{2}R_{\mathbf{cdab}}(\eta^{\mathbf{da}}%
\theta^{\mathbf{b}}-\eta^{\mathbf{db}}\theta^{\mathbf{a}})\nonumber\\
&  =-R_{\mathbf{ca}}\theta^{\mathbf{b}}=-\mathcal{R}_{\mathbf{c,}} \label{8.9}%
\end{align}
it follows that $-\theta^{\mathbf{c}}\underset{\mathtt{\mathbf{g}}}{\lrcorner
}(\theta^{\mathbf{d}}\underset{\mathtt{\mathbf{g}}}{\lrcorner}\mathcal{R}%
_{\mathbf{cd}})=\theta^{\mathbf{c}}\bullet\mathcal{R}_{c}=R$. The
$\mathcal{R}_{\mathbf{c,}}$ are called the Ricci $1$-forms.

Next we establish the following proposition.

\begin{proposition}
The Einstein-Hilbert Lagrangian can be written as
\index{Einstein-Hilbert lagrangian}
\begin{equation}
\mathcal{L}_{EH}=-d\left(  \theta^{\mathbf{a}}\wedge\underset
{\mathtt{\mathbf{g}}}{\star}d\theta_{\mathbf{a}}\right)  +\mathcal{L}_{g},
\label{8.14}%
\end{equation}
where
\begin{equation}
\mathcal{L}_{g}=-\frac{1}{2}\tau_{\mathtt{\mathbf{g}}}\theta^{\mathbf{c}%
}\underset{\mathtt{\mathbf{g}}}{\lrcorner}\theta^{\mathbf{b}}\underset
{\mathtt{\mathbf{g}}}{\lrcorner}\left(  \omega_{\mathbf{ab}}\wedge
\omega_{\mathbf{c}}^{\mathbf{a}}\right)  , \label{8.15}%
\end{equation}
is the first order Lagrangian density (first introduced by Einstein)

\begin{proof}
That $\mathcal{L}_{g}$ is the \textit{intrinsic} form of the Einstein first
order Lagrangian in the gauge defined by $\theta^{\mathbf{a}}$ is easily seen
writing $\omega_{\mathbf{b}}^{\mathbf{a}}=L_{\mathbf{bc}}^{\mathbf{a}}%
\theta^{\mathbf{c}}$. Indeed, we immediately verify using again well known
identities (see, e.g., Chapter 2 of \cite{rodoliv2006}), which give
\begin{equation}
\theta^{\mathbf{c}}\underset{\mathtt{\mathbf{g}}}{\lrcorner}\theta
^{\mathbf{b}}\underset{\mathtt{\mathbf{g}}}{\lrcorner}\left(  \omega
_{\mathbf{ac}}\wedge\omega_{\mathbf{b}}^{\mathbf{a}}\right)  =\eta
^{\mathbf{bk}}\left(  L_{\mathbf{kc}}^{\mathbf{d}}L_{\mathbf{db}}^{\mathbf{c}%
}-L_{\mathbf{dc}}^{\mathbf{d}}L_{\mathbf{kb}}^{\mathbf{c}}\right)  .
\label{8.17}%
\end{equation}
To prove that $\mathcal{L}_{EH}$ can be written as in Eq.(\ref{8.14}) we start
using Cartan's second structure equation to write Eq.(\ref{8.7}) as:%
\begin{align}
\mathcal{L}_{EH}  &  =\frac{1}{2}d\omega_{\mathbf{ab}}\wedge\underset
{\mathtt{\mathbf{g}}}{\star}(\theta^{\mathbf{a}}\wedge\theta^{\mathbf{b}%
})+\frac{1}{2}\omega_{\mathbf{ac}}\wedge\omega_{\mathbf{b}}^{\mathbf{c}}%
\wedge\underset{\mathtt{\mathbf{g}}}{\star}(\theta^{\mathbf{a}}\wedge
\theta^{\mathbf{b}})\nonumber\\
&  =\frac{1}{2}d[\omega_{\mathbf{ab}}\wedge\underset{\mathtt{\mathbf{g}}%
}{\star}(\theta^{\mathbf{a}}\wedge\theta^{\mathbf{b}})]+\frac{1}{2}%
\omega_{\mathbf{ab}}\wedge\underset{\mathtt{\mathbf{g}}}{\star}d(\theta
^{\mathbf{a}}\wedge\theta^{\mathbf{b}})+\frac{1}{2}\omega_{\mathbf{ac}}%
\wedge\omega_{\mathbf{b}}^{\mathbf{c}}\wedge\underset{\mathtt{\mathbf{g}}%
}{\star}(\theta^{\mathbf{a}}\wedge\theta^{\mathbf{b}})\nonumber\\
&  =\frac{1}{2}d[\omega_{\mathbf{ab}}\wedge\underset{\mathtt{\mathbf{g}}%
}{\star}(\theta^{\mathbf{a}}\wedge\theta^{\mathbf{b}})]-\frac{1}{2}%
\omega_{\mathbf{ab}}\wedge\omega_{\mathbf{c}}^{\mathbf{a}}\wedge
\underset{\mathtt{\mathbf{g}}}{\star}(\theta^{\mathbf{c}}\wedge\theta
^{\mathbf{b}}). \label{8.7new}%
\end{align}
Next, using again well known identities (see ,e.g., Chapter 2 of
\cite{rodoliv2006}) we get (recall that $\omega^{\mathbf{cd}}=-\omega
^{\mathbf{dc}}$)
\begin{align}
(\theta^{\mathbf{c}}\wedge\theta^{\mathbf{d}})\underset{\mathtt{\mathbf{g}}%
}{\wedge\star}\omega_{\mathbf{cd}}  &  =-\underset{\mathtt{\mathbf{g}}}{\star
}[\omega^{\mathbf{cd}}\underset{\mathtt{\mathbf{g}}}{\lrcorner}(\theta
_{\mathbf{c}}\wedge\theta_{\mathbf{d}})]\nonumber\\
&  =\underset{\mathtt{\mathbf{g}}}{\star}[(\omega^{\mathbf{cd}}\underset
{\mathtt{\mathbf{g}}}{\cdot}\theta_{\mathbf{d}})\theta_{\mathbf{c}}%
-(\omega^{\mathbf{cd}}\underset{\mathtt{\mathbf{g}}}{\cdot}\theta_{\mathbf{c}%
})\theta_{\mathbf{d}}]=2\underset{\mathtt{\mathbf{g}}}{\star}[(\omega
^{\mathbf{cd}}\underset{\mathtt{\mathbf{g}}}{\cdot}\theta_{\mathbf{d}}%
)\theta_{\mathbf{c}}],
\end{align}
and from Cartan's first structure equation we have
\end{proof}
\end{proposition}

\begin{proof}%
\begin{align}
\theta^{\mathbf{a}}\wedge\underset{\mathtt{\mathbf{g}}}{\star}d\theta
_{\mathbf{a}}  &  =\theta^{\mathbf{a}}\wedge\underset{\mathtt{\mathbf{g}}%
}{\star}(\omega_{\mathbf{ba}}\wedge\theta^{\mathbf{b}})=-\underset
{\mathtt{\mathbf{g}}}{\star}[\theta_{\mathbf{a}}\underset{\mathtt{\mathbf{g}}%
}{\lrcorner}(\omega^{\mathbf{ba}}\wedge\theta_{\mathbf{b}})]\nonumber\\
&  =-\underset{\mathtt{\mathbf{g}}}{\star\lbrack}(\theta_{\mathbf{a}}%
\underset{\mathtt{\mathbf{g}}}{\cdot}\omega^{\mathbf{ba}})\theta_{\mathbf{b}%
}]=-\underset{\mathtt{\mathbf{g}}}{\star}[(\theta^{\mathbf{a}}\underset
{\mathtt{\mathbf{g}}}{\cdot}\omega_{\mathbf{ba}})\theta_{\mathbf{b}}],
\end{align}
from where it follows that%
\begin{equation}
\frac{1}{2}d[(\theta^{\mathbf{c}}\wedge\theta^{\mathbf{d}})\wedge
\underset{\mathtt{\mathbf{g}}}{\star}\omega_{\mathbf{cd}}]=-d(\theta
^{\mathbf{a}}\wedge\underset{\mathtt{\mathbf{g}}}{\star}d\theta_{\mathbf{a}%
}]).
\end{equation}

On the other hand the second term in the last line of Eq.(\ref{8.7new}) can be
written as%
\begin{align*}
&  \frac{1}{2}\omega_{\mathbf{ab}}\wedge\omega_{\mathbf{c}}^{\mathbf{a}}%
\wedge\underset{\mathtt{\mathbf{g}}}{\star}(\theta^{\mathbf{c}}\wedge
\theta^{\mathbf{b}})\\
&  =-\frac{1}{2}\underset{\mathtt{\mathbf{g}}}{\star}[(\theta^{\mathbf{b}%
}\underset{\mathtt{\mathbf{g}}}{\cdot}\omega_{\mathbf{ab}})(\theta
^{\mathbf{c}}\underset{\mathtt{\mathbf{g}}}{\cdot}\omega_{\mathbf{c}%
}^{\mathbf{a}})-(\theta^{\mathbf{b}}\underset{\mathtt{\mathbf{g}}}{\cdot
}\omega_{\mathbf{c}}^{\mathbf{a}})(\theta^{\mathbf{c}}\underset
{\mathtt{\mathbf{g}}}{\cdot}\omega_{\mathbf{ab}})].
\end{align*}
Now,
\begin{align*}
&  (\theta^{\mathbf{b}}\underset{\mathtt{\mathbf{g}}}{\cdot}\omega
_{\mathbf{ab}})(\theta^{\mathbf{c}}\underset{\mathtt{\mathbf{g}}}{\cdot}%
\omega_{\mathbf{c}}^{\mathbf{a}})\\
&  =\omega_{\mathbf{ab}}\underset{\mathtt{\mathbf{g}}}{\cdot}[(\theta
^{\mathbf{c}}\underset{\mathtt{\mathbf{g}}}{\cdot}\omega_{\mathbf{c}%
}^{\mathbf{a}})\theta^{\mathbf{b}}]\\
&  =\omega_{\mathbf{ab}}\underset{\mathtt{\mathbf{g}}}{\cdot}[\theta
^{\mathbf{c}}\underset{\mathtt{\mathbf{g}}}{\lrcorner}(\omega_{\mathbf{c}%
}^{\mathbf{a}}\wedge\theta^{\mathbf{b}})+\omega^{\mathbf{ab}}]\\
&  =(\omega_{\mathbf{ab}}\wedge\theta^{\mathbf{c}})\underset
{\mathtt{\mathbf{g}}}{\lrcorner}(\omega_{\mathbf{c}}^{\mathbf{a}}\wedge
\theta^{\mathbf{b}})+\omega_{\mathbf{cd}}\underset{\mathtt{\mathbf{g}}}{\cdot
}\omega^{\mathbf{cd}}]
\end{align*}
and taking into account that $d\theta^{\mathbf{a}}=-\omega_{\mathbf{b}%
}^{\mathbf{a}}\wedge\theta^{\mathbf{b}}$ , $d\underset{\mathtt{\mathbf{g}}%
}{\star}\theta^{\mathbf{a}}=-\omega_{\mathbf{b}}^{\mathbf{a}}\underset
{\mathtt{\mathbf{g}}}{\cdot}\theta^{\mathbf{b}}$ and that $\delta
\theta_{\mathbf{a}}=-\underset{\mathtt{\mathbf{g}}}{\star}^{-1}d\underset
{\mathtt{\mathbf{g}}}{\star}\theta_{\mathbf{a}}$ we have%

\begin{equation}
\frac{1}{2}\omega_{\mathbf{ab}}\wedge\omega_{\mathbf{c}}^{\mathbf{a}}%
\wedge\underset{\mathtt{\mathbf{g}}}{\star}(\theta^{\mathbf{c}}\wedge
\theta^{\mathbf{b}})=\frac{1}{2}[-d\theta^{\mathbf{a}}\wedge\underset
{\mathtt{\mathbf{g}}}{\star}d\theta_{\mathbf{a}}-\underset{\mathtt{\mathbf{g}%
}}{\delta}\theta^{\mathbf{a}}\wedge\underset{\mathtt{\mathbf{g}}}{\star
}\underset{\mathtt{\mathbf{g}}}{\delta}\theta_{\mathbf{a}}+\omega
_{\mathbf{cd}}\wedge\underset{\mathtt{\mathbf{g}}}{\star}\omega^{\mathbf{cd}}]
\end{equation}

Next, using Eq.(\ref{8.19}) the last term in the last equation can be written
as%
\[
\frac{1}{2}\omega_{\mathbf{cd}}\wedge\underset{\mathtt{\mathbf{g}}}{\star
}\omega^{\mathbf{cd}}=d\theta^{\mathbf{a}}\wedge\underset{\mathtt{\mathbf{g}}%
}{\star}d\theta_{\mathbf{a}}-\frac{1}{4}\left(  d\theta^{\mathbf{a}}%
\wedge\theta_{\mathbf{a}}\right)  \wedge\underset{\mathtt{\mathbf{g}}}{\star
}\left(  d\theta^{\mathbf{b}}\wedge\theta_{\mathbf{b}}\right)
\]
and we finally get%
\begin{equation}
\mathcal{L}_{g}=-\frac{1}{2}d\theta^{\mathbf{a}}\wedge\underset
{\mathtt{\mathbf{g}}}{\star}d\theta_{\mathbf{a}}+\frac{1}{2}\underset
{\mathtt{\mathbf{g}}}{\delta}\theta^{\mathbf{a}}\wedge\underset
{\mathtt{\mathbf{g}}}{\star}\underset{\mathtt{\mathbf{g}}}{\delta}%
\theta_{\mathbf{a}}+\frac{1}{4}\left(  d\theta^{\mathbf{a}}\wedge
\theta_{\mathbf{a}}\right)  \wedge\underset{\mathtt{\mathbf{g}}}{\star}\left(
d\theta^{\mathbf{b}}\wedge\theta_{\mathbf{b}}\right)  , \label{8.20}%
\end{equation}
and the proposition is proved.
\end{proof}

Now, \ since
\[
d(\theta^{\mathbf{a}}\wedge\underset{\mathtt{\mathbf{g}}}{\star}%
d\theta_{\mathbf{a}})=\mathtt{h}^{\ast}d(\mathfrak{g}^{\mathbf{a}}%
\wedge\underset{\eta}{\star}d\mathfrak{g}_{\mathbf{a}})
\]

we see taking into account Eq.(\ref{8.1biss}) that\footnote{This result can be
used to show, as stated in the begining of the article that $\mathcal{L}%
_{g}^{M}$ is invariant under local Lorentz transformations. Indeed, we have
that $\mathcal{L}_{g}^{M}=d(\mathtt{h}^{\ast-1}(\mathfrak{g}^{\mathbf{a}%
}\wedge\underset{\mathtt{\eta}}{\star}d\mathfrak{g}_{\mathbf{a}}%
))+R\tau_{\mathtt{\eta}}$, which is manifestly invariant under the local
action of the Lorentz group, since $R$ is a scalar function and $\tau
_{\mathtt{\eta}}^{\prime}=\mathfrak{g}^{\prime0}\wedge\mathfrak{g}^{\prime
1}\wedge\mathfrak{g}^{\prime2}\wedge\mathfrak{g}^{\prime3}=u\mathfrak{g}%
^{0}u^{-1}u\mathfrak{g}^{1}u^{-1}u\mathfrak{g}^{2}u^{-1}u\mathfrak{g}%
^{3}u^{-1}=$ $\ u\tau_{\mathtt{\eta}}u^{-1}=\tau_{\mathtt{\eta}}%
=\mathfrak{g}^{0}\wedge\mathfrak{g}^{1}\wedge\mathfrak{g}^{2}\wedge
\mathfrak{g}^{3}$, for any $u\in\sec\mathrm{Spin}_{1,3}^{e}(M,\mathtt{\eta
})\hookrightarrow\sec\mathcal{C}\ell(M,\mathtt{\eta})$.}
\begin{equation}
\mathcal{L}_{EH}=-d(\theta^{\mathbf{a}}\wedge\underset{\mathtt{\mathbf{g}}%
}{\star}d\theta_{\mathbf{a}})+\mathcal{L}_{g}=\mathtt{h}^{\ast}%
[-d(\mathfrak{g}^{\mathbf{a}}\wedge\underset{\eta}{\star}d\mathfrak{g}%
_{\mathbf{a}})+\mathcal{L}_{g}^{M}]=\mathtt{h}^{\ast}\mathcal{L}_{EH}^{M}
\label{8.20a}%
\end{equation}

Since the variational principle $%
\mbox{\boldmath{$\delta$}}%
{\displaystyle\int}
\mathcal{L}_{EH}=0$ implies $%
\mbox{\boldmath{$\delta$}}%
{\displaystyle\int}
\mathtt{h}^{\ast-1}\mathcal{L}_{EH}=0$ and thus $%
\mbox{\boldmath{$\delta$}}%
{\displaystyle\int}
\mathcal{L}_{EH}^{M}$ $=0$ we just proved that as stated $\mathcal{L}_{EH}$
and $\mathcal{L}_{EH}^{M}$ are indeed equivalent.

\section{Maxwell like Form of the Gravitational Equations}

Call $\mathfrak{F}^{\mathbf{a}}=d\mathfrak{g}^{\mathbf{a}}$. Let us verify
that the $\mathfrak{F}^{\mathbf{a}}$ satisfy Maxwell (like) equations, i.e.,
when the energy-momentum tensor of matter the fields is non zero we have in
general
\begin{equation}
d\mathfrak{F}^{\mathbf{a}}=0,\text{ }\underset{\eta}{\delta}\mathfrak{F}%
^{\mathbf{a}}=J_{\mathbf{m}}^{\mathbf{a}}, \label{8.27h}%
\end{equation}
where $J_{\mathbf{m}}^{\mathbf{a}}$ is an appropriate current which we now determine.

Since $\underset{\eta}{\delta}\mathfrak{F}^{\mathbf{a}}=\mathtt{h}^{\ast
-1}\underset{\mathtt{\mathbf{g}}}{\delta}d\theta^{\mathbf{a}}$ we can write
remembering the definition of the Hodge Laplacian
\begin{equation}
-\mathtt{h}^{\ast-1}\underset{\mathtt{\mathbf{g}}}{\delta d}\theta
^{\mathbf{a}}=-\mathtt{h}^{\ast-1}(\underset{\mathtt{\mathbf{g}}}{\delta
}d\theta^{\mathbf{a}}+\text{ }\underset{\mathtt{\mathbf{g}}}{d\delta}%
\theta^{\mathbf{a}})+\mathtt{h}^{\ast-1}\underset{\mathtt{\mathbf{g}}}%
{d\delta}\theta^{\mathbf{a}}=\mathtt{h}^{\ast-1}\Diamond\theta^{\mathbf{a}%
}+\mathtt{h}^{\ast-1}\underset{\mathtt{\mathbf{g}}}{d\delta}\theta
^{\mathbf{a}}%
\end{equation}
Now, recall from \cite{rodqui,rodoliv2006} that the Hodge Laplacian can be
written as the square of the Dirac operator ${\mbox{\boldmath$\partial$}}%
=\theta^{\mathbf{c}}\nabla_{\mathbf{e}_{\mathbf{c}}}$ associated to $\nabla$,
the Levi-Civita connection of $\mathtt{\mathbf{g,}}$i.e.\textbf{, }for any
$K\in\sec%
{\displaystyle\bigwedge\nolimits^{p}}
T^{\ast}M\hookrightarrow\mathcal{C\ell(}M,\mathtt{g}),$%
\[
\Diamond K=-{\mbox{\boldmath$\partial$}}^{2}%
K=({\mbox{\boldmath$\partial$}\wedge\mbox{\boldmath$\partial$})}%
K+({\mbox{\boldmath$\partial$}}\underset{\mathtt{\mathbf{g}}}{{\cdot}%
}{\mbox{\boldmath$\partial$})}K,
\]
where ${\mbox{\boldmath$\partial$}\wedge\mbox{\boldmath$\partial$}}$ is called
the Ricci operator and ${\mbox{\boldmath$\partial$}}\underset
{\mathtt{\mathbf{g}}}{{\cdot}}{\mbox{\boldmath$\partial$}=\square}$ is the
covariant D' Alembertian. We have\footnote{Eq.(\ref{hodge}) permits the
comparison of the Lagrange multiplier $\lambda(x)$ appearing in the theory
presented in \cite{itina,itin,kaniel} (where the field equations are
\ $\Diamond\theta^{\mathbf{a}}+\lambda(x)\theta^{\mathbf{a}}=0$) with its
value in General Relativity.} \cite{rodqui,rodoliv2006}
\begin{align}
\Diamond\theta^{\mathbf{a}} &  =({\mbox{\boldmath$\partial$}\wedge
\mbox{\boldmath$\partial$})}\theta^{\mathbf{a}}+\square\theta^{\mathbf{a}%
}\nonumber\\
&  =\mathcal{R}^{\mathbf{a}}+\square\theta^{\mathbf{a}}\nonumber\\
&  =\mathcal{T}^{\mathbf{a}}-\frac{1}{2}\mathcal{T}\theta^{\mathbf{a}}%
+\square\theta^{\mathbf{a}},\label{hodge}%
\end{align}
where $\mathcal{R}^{\mathbf{a}}$ are the Ricci $1$-forms (Eq.(\ref{8.9})).
Taking into account also that,%
\begin{equation}
\square\theta^{\mathbf{c}}=-\frac{1}{2}\eta^{\mathbf{ab}}M_{\mathbf{d\hspace
{0.2cm}ab}}^{\hspace{0.2cm}\mathbf{c}}\theta^{\mathbf{d}},
\end{equation}
where the tensor field $M_{\mathbf{d\hspace{0.2cm}ab}}^{\hspace{0.2cm}%
\mathbf{c}}$ is given by \cite{rodquin,rodoliv2006}
\begin{equation}
M_{\mathbf{d\hspace{0.2cm}ab}}^{\hspace{0.2cm}\mathbf{c}}=\mathbf{e}%
_{\mathbf{a}}(L_{\mathbf{bd}}^{\mathbf{c}})+\mathbf{e}_{\mathbf{b}%
}(L_{\mathbf{ad}}^{\mathbf{c}})-L_{\mathbf{ak}}^{\mathbf{c}}L_{\mathbf{bd}%
}^{\mathbf{k}}-L_{\mathbf{bk}}^{\mathbf{c}}L_{\mathbf{ad}}^{\mathbf{k}%
}-(L_{\mathbf{ab}}^{\mathbf{k}}.+L_{\mathbf{ba}}^{\mathbf{k}})L_{\mathbf{kd}%
}^{\mathbf{c}},
\end{equation}
with $\nabla_{\mathbf{e}^{\mathbf{a}}}\theta^{\mathbf{b}}=-L_{\mathbf{ac}%
}^{\mathbf{b}}\theta^{\mathbf{c}}$ and putting moreover,
\[
\mathtt{h}^{\ast-1}\mathcal{T}^{\mathbf{a}}=\mathcal{T}_{M}^{\mathbf{a}},
\]
we finally have
\begin{equation}
\underset{\eta}{\delta}\mathfrak{F}^{\mathbf{a}}=-\mathfrak{J}^{\mathbf{a}%
},\label{xxx}%
\end{equation}
with%
\begin{equation}
\mathfrak{J}^{\mathbf{a}}=\mathcal{T}_{M}^{\mathbf{a}}-\frac{1}{2}%
\mathcal{T}\mathfrak{g}^{\mathbf{a}}+\frac{1}{2}\eta^{\mathbf{lb}%
}M_{\mathbf{d\hspace{0.2cm}lb}}^{\hspace{0.2cm}\mathbf{a}}\mathfrak{g}%
^{\mathbf{d}}-d\underset{\eta}{\delta}\mathfrak{g}^{\mathbf{a}}\label{current}%
\end{equation}

The currents $\mathfrak{J}^{\mathbf{a}}$ are conserved, i.e., $\underset{\eta
}{\delta}$ $\mathfrak{J}^{\mathbf{a}}=0$ and express the energy-momentum
conservation law for the system composed by the gravitational and matter
fields. In particular, from Eq.(\ref{current}) we are tempted to call%
\begin{equation}
t_{g}^{\mathbf{a}}=\frac{1}{2}\eta^{\mathbf{lb}}M_{\mathbf{d\hspace{0.2cm}lb}%
}^{\hspace{0.2cm}\mathbf{a}}\mathfrak{g}^{\mathbf{d}}-d\underset{\eta}{\delta
}\mathfrak{g}^{\mathbf{a}} \label{current 2}%
\end{equation}
the true energy momentum $1$-forms of the gravitational field and $-\frac
{1}{2}\mathcal{T}\mathfrak{g}^{\mathbf{a}}$ the interaction energy-momentum
$1$-forms. This will be investigate further in another publication.

\begin{remark}
Note that $\mathfrak{F}^{\mathbf{a}}=d\mathfrak{g}^{\mathbf{a}}=\mathtt{h}%
^{\ast-1}d\theta^{\mathbf{a}}$. In the teleparallel equivalent of GRT (see
below) the Lorentzian manifold $(M,\mathtt{\mathbf{g}})$ is equipped with a
teleparallel connection such that the torsion $2$-forms are $\Theta
^{\mathbf{a}}=d\theta^{\mathbf{a}}$. Then, \emph{Eq.(\ref{xxx})} can be used
to write an equation for $\underset{\mathtt{\mathbf{g}}}{\delta}%
\Theta^{\mathbf{a}}$, which the reader may find without difficulties.
\end{remark}

\section{Genuine Energy-Momentum Conservation Law}

Once we showed that it is possible to express Einstein's gravitational
equations as the equations of physical fields $\mathfrak{g}^{\mathbf{a}}$ in
Minkowski spacetime we look for the well known\footnote{See,
e.g.,\cite{rodoliv2006,thiwal}.} form of Einstein's equations in terms of the
superpotentials $\mathcal{S}^{\mathbf{a}}$, and which we write here as%
\begin{equation}
-d\underset{\mathtt{\mathbf{g}}}{\star}\mathcal{S}^{\mathbf{a}}=\text{
}\underset{\mathtt{\mathbf{g}}}{\star}\mathcal{T}^{\mathbf{a}}+\text{
}\underset{\mathtt{\mathbf{g}}}{\star}t^{\mathbf{a}},\label{8.27}%
\end{equation}
with
\begin{align}
\underset{\mathtt{\mathbf{g}}}{\star}t_{\mathbf{\ }}^{\mathbf{c}} &
=\frac{\partial\mathcal{L}_{g}}{\partial\theta_{a}}=-\frac{1}{2}%
\mbox{\boldmath{$\omega$}}%
_{\mathbf{ab}}\wedge\lbrack%
\mbox{\boldmath{$\omega$}}%
_{\mathbf{d}}^{\mathbf{c}}\wedge\underset{\mathtt{\mathbf{g}}}{\star}%
(\theta^{\mathbf{a}}\wedge\theta^{\mathbf{b}}\wedge\theta^{\mathbf{d}})+%
\mbox{\boldmath{$\omega$}}%
_{\mathbf{d}}^{\mathbf{b}}\wedge\underset{\mathtt{\mathbf{g}}}{\star}%
(\theta^{\mathbf{a}}\wedge\theta^{\mathbf{d}}\wedge\theta^{\mathbf{c}}%
)]\in\sec\bigwedge\nolimits^{3}T^{\ast}M\hookrightarrow\mathcal{C\ell}\left(
T^{\ast}M,\mathtt{g}\right)  \nonumber\\
\underset{\mathtt{\mathbf{g}}}{\star}\mathcal{S}^{\mathbf{c}} &
=\frac{\partial\mathcal{L}_{g}}{\partial d\theta_{a}}=\frac{1}{2}%
\mbox{\boldmath{$\omega$}}%
_{\mathbf{ab}}\wedge\underset{\mathtt{\mathbf{g}}}{\star}(\theta^{\mathbf{a}%
}\wedge\theta^{\mathbf{b}}\wedge\theta^{\mathbf{c}})\in\sec\bigwedge
\nolimits^{2}T^{\ast}M\hookrightarrow\mathcal{C\ell}\left(  T^{\ast
}M,\mathtt{g}\right)  ,\label{7.10.17}%
\end{align}
and where $%
\mbox{\boldmath{$\omega$}}%
_{\mathbf{ab}}$ is given by Eq.(\ref{8.19})

As discussed,e.g., in \cite{rodsouza} Eq.(\ref{8.27}) does not express any
trustful energy-momentum conservation law in a general Lorentzian spacetime.
However, it express a trustful energy-momentum conservation law in Minkowski
spacetime, since it is equivalent (as the reader may verify) to%

\begin{equation}
-dh^{-1}\star h\mathcal{S}^{\mathbf{a}}=h^{-1}\star h\mathcal{T}^{\mathbf{a}%
}+\text{ }h^{-1}\star ht^{\mathbf{a}}, \label{8.28}%
\end{equation}
where in Eq.(\ref{8.28}),%
\begin{equation}
\star\label{8.29}%
\end{equation}
is the Hodge dual relative to the Minkowski metric $%
\mbox{\boldmath{$\eta$}}%
=\eta_{\mathbf{ab}}\vartheta^{\mathbf{a}}\otimes\vartheta^{\mathbf{b}}$.

\subsection{Mass of the Graviton}

In the Lagrangian given by Eq.(\ref{8.1bis}) the mass of the graviton is
supposed to be zero. A non null mass $m$ requires an extra term in the
Lagrangian. As an example, consider the Lagrangian density%
\begin{equation}
\mathcal{L}_{g}^{\prime}=-\frac{1}{2}d\theta^{\mathbf{a}}\wedge\underset
{\mathtt{\mathbf{g}}}{\star}d\theta_{\mathbf{a}}+\frac{1}{2}\underset
{\mathtt{\mathbf{g}}}{\delta}\theta^{\mathbf{a}}\wedge\underset
{\mathtt{\mathbf{g}}}{\star}\underset{\mathtt{\mathbf{g}}}{\delta}%
\theta_{\mathbf{a}}+\frac{1}{4}\left(  d\theta^{\mathbf{a}}\wedge
\theta_{\mathbf{a}}\right)  \wedge\underset{\mathtt{\mathbf{g}}}{\star}\left(
d\theta^{\mathbf{b}}\wedge\theta_{\mathbf{b}}\right)  +\frac{1}{2}m^{2}%
\theta_{\mathbf{a}}\wedge\underset{\mathtt{\mathbf{g}}}{\star}\theta
^{\mathbf{a}} \label{war1}%
\end{equation}

With the extra term the equations for the gravitational field, for the
$\mathcal{S}^{\mathbf{a}}$ result in%
\begin{equation}
-d\underset{\mathtt{\mathbf{g}}}{\star}\mathcal{S}^{\mathbf{a}}=\underset
{\mathtt{\mathbf{g}}}{\star}\mathcal{T}^{\mathbf{a}}+\underset
{\mathtt{\mathbf{g}}}{\text{ }\star}t^{\mathbf{a}}+m^{2}\underset
{\mathtt{\mathbf{g}}}{\star}\theta^{\mathbf{a}}, \label{war2}%
\end{equation}
from where we get
\begin{equation}
\underset{\mathtt{\mathbf{g}}}{\delta}(\mathcal{T}^{\mathbf{a}}+t^{\mathbf{a}%
})=-m^{2}\underset{\mathtt{\mathbf{g}}}{\delta}\theta^{\mathbf{a}}
\label{war3}%
\end{equation}

If we impose the gauge $\underset{\mathtt{\mathbf{g}}}{\delta}\theta
^{\mathbf{a}}=0$, which is analogous to the Lorenz gauge in electrodynamics,
Eq.(\ref{war3}) becomes
\begin{equation}
\underset{\mathtt{\mathbf{g}}}{\delta}(\mathcal{T}^{\mathbf{a}}+t^{\mathbf{a}%
})=0, \label{war4}%
\end{equation}
which is the same equation valid in the case $m=0$!

There are other possibilities of having a non null graviton mass, as, e.g., in
Logunov's theory \cite{logunov1,logunov2}, which we do not discuss
here\footnote{We only observe that Lagrangian density of Logunov's theory when
written in terms of differential forms is not a very elegant expression.}.

\section{ Teleparallel Geometry}

Before closing this paper we briefly recall our discussion in \cite{roroqui}
where it was observed that recently some people \cite{deandrade} think to have
find a valid way of formulating a genuine energy-momentum conservation law in
a theory that is claimed to be (and indeed, it is) equivalent to general
relativity. In that theory, the so-called \textit{teleparallel} equivalent of
General Relativity theory \cite{maluf}, spacetime is teleparallel (or
Weintzb\"{o}ck), i.e., has a \ metric compatible connection with non zero
torsion and with null curvature\footnote{In fact, formulation of teleparallel
equivalence of General Relativity is a subject with a old history. See, e.g.,
\cite{haya}.}. We showed in \cite{roroqui} that the claim of \cite{deandrade}
must be qualified. Here, our objective is only to show that the teleparallel
theory is a possible trivial interpretation of our formalism. Indeed, the
structure of the teleparallel equivalent of GRT as formulated, e.g., by
\cite{maluf} or \cite{deandrade} consists in nothing more than a trivial
introduction of: (i) \ a bilinear form (a deformed metric tensor)
\texttt{$\mathbf{g}$ }$=\eta_{\mathbf{ab}}\theta^{\mathbf{a}}\otimes
\theta^{\mathbf{b}}$ and (ii) a teleparallel connection in the manifold
$M\simeq\mathbb{R}^{4}$ of Minkowski spacetime structure. Indeed, taking
advantage of the the discussion of the previous sections, we can present that
theory with a cosmological constant term as follows. Start with $\mathcal{L}%
_{g}^{\prime}$ (Eq.(\ref{war1})) and write it (after some algebraic
manipulations) as%
\begin{align}
\mathcal{L}_{g}^{\prime}  &  =-\frac{1}{2}d\theta^{\mathbf{a}}\wedge
\underset{\mathtt{\mathbf{g}}}{\star}\left[  d\theta_{\mathbf{a}}%
-\theta_{\mathbf{a}}\wedge(\theta_{\mathbf{b}}\underset{\mathtt{\mathbf{g}}%
}{\lrcorner}d\theta_{\mathbf{b}})+\frac{1}{2}\underset{\mathtt{\mathbf{g}}%
}{\star}\left(  \theta_{\mathbf{a}}\wedge\star(d\theta^{\mathbf{b}}%
\wedge\theta_{\mathbf{b}})\right)  \right]  +\frac{1}{2}m^{2}\theta
_{\mathbf{a}}\wedge\underset{\mathtt{\mathbf{g}}}{\star}\theta^{\mathbf{a}%
}\nonumber\\
&  =-\frac{1}{2}d\theta^{\mathbf{a}}\wedge\underset{\mathtt{\mathbf{g}}}%
{\star}(^{(1)}d\theta_{\mathbf{a}}-2^{(2)}d\theta_{\mathbf{a}}-\frac{1}%
{2}\text{ }^{(3)}d\theta_{\mathbf{a}})+\frac{1}{2}m^{2}\theta_{\mathbf{a}%
}\wedge\underset{\mathtt{\mathbf{g}}}{\star}\theta^{\mathbf{a}}, \label{tele1}%
\end{align}
where
\begin{align}
d\theta^{\mathbf{a}}  &  =\text{ }^{{\small (1)}}d\theta^{\mathbf{a}}+\text{
}^{{\small (2)}}d\theta^{\mathbf{a}}+\text{ }^{{\small (3)}}d\theta
^{\mathbf{a}},\nonumber\\
^{{\small (1)}}d\theta^{\mathbf{a}}  &  =d\theta^{\mathbf{a}}-\text{
}^{{\small (2)}}d\theta^{\mathbf{a}}-\text{ }^{{\small (3)}}d\theta
^{\mathbf{a}},\nonumber\\
^{{\small (2)}}d\theta^{\mathbf{a}}  &  =\frac{1}{3}\theta^{\mathbf{b}}%
\wedge(\theta_{\mathbf{b}}\underset{\mathtt{\mathbf{g}}}{\lrcorner}%
d\theta_{\mathbf{b}}),\nonumber\\
^{{\small (3)}}d\theta^{\mathbf{a}}  &  =-\frac{1}{3}\star\left(
\theta^{\mathbf{b}}\wedge\star(d\theta^{\mathbf{b}}\wedge\theta_{\mathbf{b}%
})\right)  . \label{tele2}%
\end{align}
Next introduce a teleparallel connection by declaring that the cobasis
$\{\theta^{\mathbf{a}}\}$ fixes the parallelism, i.e., we define the torsion
$2$-forms by
\begin{equation}
\Theta^{\mathbf{a}}:=d\theta^{\mathbf{a}}, \label{tele3}%
\end{equation}

and $\mathcal{L}_{g}$ becomes%
\begin{equation}
\mathcal{L}_{g}=-\frac{1}{2}\Theta^{\mathbf{a}}\wedge\underset
{\mathtt{\mathbf{g}}}{\star}\left(  ^{{\small (1)}}\Theta^{\mathbf{a}%
}-2^{({\small 2)}}\Theta^{\mathbf{a}}-\frac{1}{2}\text{ }^{{\small (3)}}%
\Theta^{\mathbf{a}}\text{ }\right)  +\frac{1}{2}m^{2}\theta_{\mathbf{a}}%
\wedge\underset{\mathtt{\mathbf{g}}}{\star}\theta^{\mathbf{a}}, \label{tele4}%
\end{equation}
where $^{{\small (1)}}\Theta^{\mathbf{a}}=^{{\small (1)}}d\theta^{\mathbf{a}}%
$, $^{({\small 2)}}\Theta^{\mathbf{a}}=^{{\small (31)}}d\theta^{\mathbf{a}}$
and $^{{\small (3)}}\Theta^{\mathbf{a}}=^{{\small (31)}}d\theta^{\mathbf{a}}$,
called \ \textit{tractor} (four components), \textit{axitor} (four components)
and \textit{tentor} (sixteen components) are the irreducible components of the
tensor torsion under the action of \textrm{SO}$_{1,3}^{e}$.\medskip

\section{Conclusion}

In\ the writing of this paper we have been motivated, first by the desire of
having genuine energy-momentum and angular momentum conservations laws for the
gravitational and matter fields, and second by some thoughts of Kiehn
\cite{kiehn} about the physical vacuum. We thus produced (using the Clifford
bundle formalism) a theory where the gravitational field represented by
$\mathfrak{F}^{\mathbf{a}}=d\mathfrak{g}^{\mathbf{a}}$ (which are physical
fields in the Faraday sense, living in Minkowski spacetime, .like the
electromagnetic field) satisfy Maxwell like equations, $d\mathfrak{F}%
^{\mathbf{a}}=0$, $\underset{\mathtt{\mathbf{\eta}}}{\delta}\mathfrak{F}%
^{\mathbf{a}}=\mathfrak{J}^{\mathbf{a}}$, where the currents $\mathfrak{J}%
^{\mathbf{a}}$ are given by Eq.(\ref{current}). Moreover, we showed that when
the graviton mass is zero, the gravitational field can be interpreted as
creating: \textbf{(i)} an effective Lorentzian geometry where probe particles
and probe fields move, or \textbf{(ii) }an effective teleparallel geometry
where probe particles and probe fields move. In such theory there are, of
course no exotic topologies, black-holes\footnote{There are several
interesting articles criticizing the notion that black-holes are predicitons
of General Relativity on mathematical and physical grounds, as,
e.g.\cite{abrams,chapline,mottomazur,strav}. Also the \ `pasticcio' concerning
the black-hole information \ `paradox' (see, \cite{hayard,hawking}) is an
example that the foundations of General Relativity are not well understood as
some people would like us to think.}, worm-holes, no possibility for
time-machines\footnote{The possibility for time machines arises when closed
timelike curves exist in a Lorentzian manifold. Such exotic configurations, it
is \textit{said}, already appears in G\"{o}del's universe model. However, a
recent thoughtful analysis by Cooperstock and Tieu \cite{cotiu} shows that the
old claim is wrong. Authors like, e.g, Davies \cite{davies} (which are
proposing to build time machines even at home), Gott \cite{gott} and Novikov
\cite{novikov} are invited to read \cite{cotiu} and find a error in the
argument \ of that authors.}, etc., which according to our opinion are pure
science fiction objects. Eventually, many will not like the viewpoint just
presented, but we feel that many will become interested in exploiting new
ideas presented with nice Mathematics, which may be more close to the way
Nature operates.

\end{document}